\documentclass[prl,showpacs, floatfix, twocolumn]{revtex4}
\usepackage{graphicx}
\usepackage{color}
\usepackage{amsmath}
\usepackage{amssymb}
\usepackage{latexsym}
\usepackage{psfrag}

\begin{document}

\title{Edge disorder induced Anderson localization and conduction gap in
graphene nanoribbons}
\author{M. Evaldsson}
\author{I. V. Zozoulenko}
\affiliation{Solid State Electronics, Department of Science and Technology, Link\"{o}ping
University, 60174 Norrk\"{o}ping, Sweden}
\author{Hengyi Xu}
\author{T. Heinzel}
\affiliation{Condensed Matter Physics Laboratory, Heinrich-Heine-Universit\"at,
Universit\"atsstr.1, 40225 D\"usseldorf, Germany}
\date{\today}

\begin{abstract}
We study the effect of the edge disorder on the conductance of the graphene
nanoribbons (GNRs). We find that only very modest edge disorder is
sufficient to induce the conduction energy gap in the otherwise metallic
GNRs and to lift any difference in the conductance between nanoribbons of
different edge geometry. We relate the formation of the conduction gap to
the pronounced edge disorder induced Anderson-type localization which leads
to the strongly enhanced density of states at the edges, formation of
surface-like states and to blocking of conductive paths through the ribbons.
\end{abstract}

\pacs{73.63.-b, 72.10.-d, 73.22.-f, 73.23.Ad}
\maketitle

%
%
%
%

\textit{Introduction.} The discovery of single layer graphene sheets\cite%
{Novoselov2004} has generated both surprise and interest over the past few
years. Surprise because pure two-dimensional sheets were for a long time
thought to be thermodynamically unstable\cite{unstableGraphene}. Interest
because graphene shows some extraordinary properties. Its charge carriers
mimic relativistic particles and can be described by the Dirac equation\cite%
{Novoselov2004,Novoselov2005_nature, Zhang}. Furthermore it has shown a high
mobility both at room temperature and at a high degree of doping\cite%
{Schedin_nature}. The latter makes graphene nano ribbons (GNRs) a strong
candidate for building blocks in future electronic devices\cite%
{Novoselov2007_nature, Chen2007}. However, one problem GNR-electronics faces
is the absence of the energy gap which makes it difficult to control
electronic and transport properties of the graphene-based devices. This
problem can be addressed by making sufficiently narrow GNR:s and thereby
augment the energy gap. The tight-binding calculations (or solutions of the
Dirac's equation based on them) indicate that the width of the gap depends
sensitively on the geometry of edges and the width of the nanoribbons\cite%
{Fujita1996,Nakada}.

The fundamental question of band gap engineering in graphene nanoribbons has
been recently addressed in several experimental studies\cite%
{Chen2007,Han2007,Li2008_Science} whose results have been strikingly
different from the expectations based on the models for ideal GNRs. In
particular, the conductance of the GNR did not exhibit the metallic behavior
expected for the ideal zigzag ribbons. Moreover, the experiment did not show
any difference between the armchair and zigzag GNRs. It is clear that the
edges and the confinement are responsible for these observations, but no
consensus has been reached yet on the origin of this remarkable behavior.
The factors that might lead to this behavior include scattering on rough
boundaries\cite{Chen2007,Han2007,Louis2007,Li2008,Gunlycke,Querlioz2008},
imperfections on the atomic scale\cite{Chen2007}, impurity scattering\cite%
{Lherbier2008}, electron interaction and/or modification of the electronic
structure due to the edge effects\cite{Cohen,Barone2006}, and even the
Coulomb blockade effects\cite{Sols2007}. It should be stressed however that
because of computation limitations most of the reported theoretical studies
such as the calculations of the mobility edge\cite{Querlioz2008},
conductance calculations addressing the effect of the edge disorder\cite%
{Li2008,Louis2007,Gunlycke}, as well as the DFT-based electronic structure calculations%
\cite{Cohen,Barone2006} (predicting the gap opening in otherwise semiconducting ribbons)
are performed for narrow GNRs where the widths are far from the range of widths of
nanoribbons studied experimentally (such as those of Ref.\onlinecite{Han2007}). Because
any edge effect is far stronger for a narrow ribbon it is not always clear how modeling
in narrow ribbons and experiments in wide ribbons relate to each other.

In this paper we present a systematic study of the conductance of realistic
edge-disordered GNRs whose dimensions are similar to those studied
experimentally\cite{Han2007,Chen2007,Li2008_Science}.
Our calculations are in excellent qualitative agreement with all the finding
reported by Han \textit{et al.}\cite{Han2007}. We find that only very modest
edge disorder is needed to induce the energy gap in the otherwise metallic
GNRs and to lift any difference in the conductance between nanoribbons of
different edge geometry. We relate the formation of the conduction gap to
the pronounced Anderson-type localization which is induced by edge disorder
and leads to the strongly enhanced density of states at the edges and to
blocking of the conductive channels through the ribbons.

\begin{figure}[tbh]
\includegraphics[scale=1]{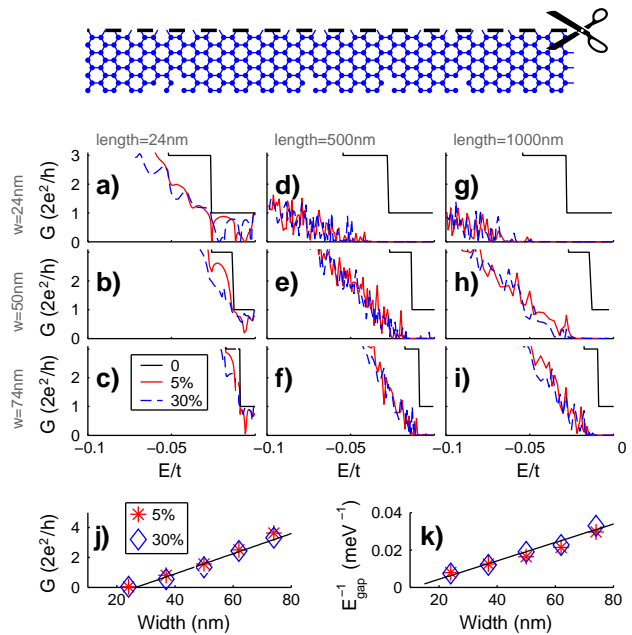}
\caption{(color online) (a)-(i) Conductance through armchair-GNR with edge
disorder and length/width as indicated in the figure. The top inset
illustrates the disordered edge with $p=5\%$. (j) Average conductance in the
energy interval $-0.061t<E<-0.049t$ versus ribbon width for $L=1\protect\mu $%
m armchair GNRs with $p$=5\% and 30\%. (k) $E_{gap}^{-1}$ versus ribbon width for the
same armchair GNRs as in (j). The solid lines in (j), (k) represent a fit as described in
the text. We define the energy gap $E_{gap}$ as the interval where $G\lesssim
10^{-3}\times 2e^2/h$ (which is consistent with the corresponding definition in
\protect\cite{Han2007}). Temperature $T=0$. } \label{fig:conductance}
\end{figure}

\textit{Model. }We describe graphene nanoribbons by the standard
tight-binding Hamiltonian on a honeycomb lattice,
\begin{equation}
H=\sum_{r}V_{r}a_{r}^{+}a_{r}-\sum_{r,r^{\prime }}t_{r,r^{\prime
}}a_{r}^{+}a_{r^{\prime }},  \label{eq:hamiltonian}
\end{equation}%
where $V_{r}$ is the external potential at the site $r$ and $t_{r,r^{\prime
}}=2.7$ eV is the overlap integral between neighboring sites $r$ and $%
r^{\prime }$. The summation of $r$ runs over the entire GNR lattice while $%
r^{\prime }$ is restricted to the sites next to $r$. We calculate the
conductance of the GNRs on the basis of the standard Landauer formalism. The
GNR is divided into a central region with the edge disorder of the length $L$
the width $W$ and connected to two semi-infinite leads (represented by ideal
ribbons of the same width $W$) from which electrons are injected. On the
edge of the GNR we model atoms missing from the lattice by setting the
appropriate hopping elements $t_{r,r^{\prime }}$ to zero. For the case of
the armchair GNRs sites are removed in the outermost (edge) row with the
probability $p$ (see inset in Fig. \ref{fig:conductance} for illustration).
For the case of the zigzag GNRs (which are less sensitive to edge disorder),
one more site (next to the already missing one on the edge) is removed in
the next row with the probability $p^{\prime }.$ Note that a particular
choice of disorder on the edge is not very important, see Fig. \ref{fig:edge}
below and the related discussion. As graphene is known to have few crystal
defects in general\cite{Schedin_nature} we do not remove sites inside the
GNRs. We also disregard the effect of capturing of H-atoms by the dangling
bonds at the edge which is shown to be of a minor importance for the ribbons
wider than a few nanometers.\cite{Cohen,Barone2006}.

The conductance is calculated on the basis of the Landauer formula, $%
G=-2e^{2}/h\int dET(E)\frac{\partial f_{FD}(E-E_{F})}{\partial E}$,where $%
f_{FD}$ is the Fermi-Dirac function. To compute the transmission coefficient
$T(E)$ we rely on our recent implementation of the recursive Green's
function technique for GNRs \cite{Xu2008}. In contrast to other existing
implementations, this method does not require self-consistent calculations
of the surface Greens function, which makes it far more efficient in
comparison to other methods and allows studying GNRs of realistic dimensions.

\textit{Results and Discussion}. Figure \ref{fig:conductance} shows the
conductance of the armchair GNRs of varying lengths ($L=24,500,1000$nm) and
widths ($W=24,50,74$nm) for two representative edge disorders $p=5\%$ and
30\%. The ribbon widths are nominally identical to those studied by Han
\textit{et al.}\cite{Han2007} (where the length was $L\sim 1\mu $m). Figure %
\ref{fig:conductance} (a) shows the conductance of the shortest and the most
narrow ribbon, $24\times 24$nm. Although no clear energy gap is present, the
conductance is strongly affected at all degrees of disorder in comparison to
the case of ideal GNRs. In wider ribbons of the same length, Fig. \ref%
{fig:conductance}(b),(c), the conductance increases more steeply which is a direct
consequence of increase of the number of propagating modes in the GNR of larger width.
For longer ribbons, $L\gtrsim 0.2\mu $m, the energy gap comparable to the energy interval
for the lowest propagating mode opens up in the conductance. Outside the energy gap the
conductance is significantly damped compared to the ribbons without disorder. Notably,
the energy gap in the vicinity of the Dirac point, and the conductance outside the Dirac
point are practically saturated for the edge disorder as low as $p=2-5\%$ and they change
linearly as $L$ increases. Qualitatively the same conductance (not shown here) is
obtained for the zigzag GNRs with the disorder strength $p=30\%,p^{\prime }=50\%$.

%
%
\begin{figure}[tbh]
\includegraphics[scale=0.32]{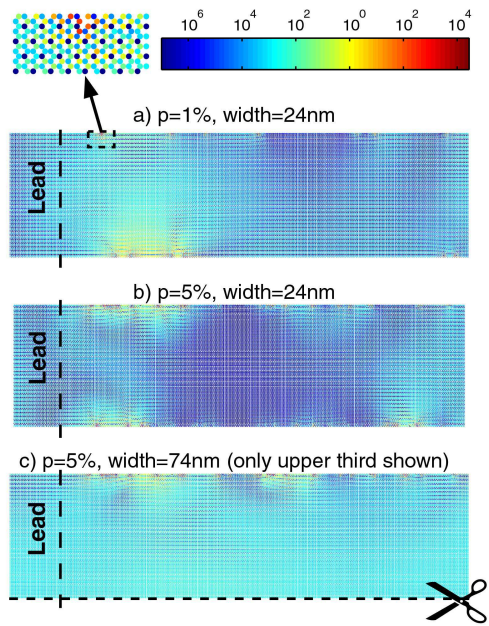}
\caption{(color online) The local density of states in a representative part
of the edge-disordered region of infinite armchair nanoribbon (with leads)
[note the logarithmic scale!]. The total width of the disordered region $%
L=150$nm, $E=-0.02t$. The defect concentration (a) $p=1\%$ (b) and (c) $%
p=5\% $. (a),(b) $W=24$nm; (c) $W=74$nm.}
\label{fig:LDOS}
\end{figure}

In order to shed a light on the origin of the conduction gap we study the
local density of states (LDOS) in the GNRs. Figure \ref{fig:LDOS} (a) shows
the LDOS in an infinite ribbon (with leads) of the width of $W=24$ nm and
the length of the disordered region $L=150$ nm (defect concentration $p=1\%)$
for the energy $E=-0.02t$. [For the shown disorder configuration and
concentration, the transmission of the GNR is $T\approx 0.1,$ which means
that there is a conductive path that allows electron to pass through the
ribbon from the left to the right lead]. The LDOS shows the Anderson-type
localization with a strongly enhanced intensity near the defects at the
ribbon edges (note the logarithmic scale of the plots!). A closer zoom
demonstrates that in the direct vicinity of the defects the magnitude of the
LDOS exceeds its value in the leads by $\sim 5-6$ orders of magnitude (see
inset on the top). The overall pattern of the LDOS shows hills (large LDOS)
and canyons (low LDOS) whose locations are clearly correlated with the
position of the disorders at the edges. With further increase of the edge
disorder concentration, a surface-like state with the enhanced density forms
over the entire edge of the ribbon. When the edge disorder concentration
increases the canyons deepen and widen and get extended over the whole width
of the ribbon blocking the conductive pathes. This is illustrated in Fig. %
\ref{fig:LDOS} (b) for a ribbon with the defect concentration $p=5\%$
(transmission $T\sim 10^{-5}\times 2e^{2}/h$ ) where such a canyon (dark
blue area) is clearly seen.

When the width of the ribbon increases the strong enhancement of the LDOS
near the edges remains practically unaffected. This is illustrated in Fig. %
\ref{fig:LDOS} (c), showing a wider ribbon of $W=74$ nm with $p=5\%$.
However the disorder induced LDOS variations do not any longer extend over
the entire width of the ribbon leaving a wide transmission path for
electrons open. This explains the absence of the conduction gap in the wider
ribbons. Note that calculated transmission in this case is $T\approx 2.5$
(with 5 propagating states in the leads).

Let us now compare quantitatively the results of our modelling to the
corresponding experimental data of Han \emph{et al.}\cite{Han2007}. The
measured conductance has been shown to scale linearly with the GNR width, $%
G=\sigma \frac{W-W^{0}}{L},$with $\sigma \approx 1.7$ mS and $W^{0}=15$ nm.
Our fit gives the same linear dependence with close values of $\sigma
\approx 5.2$ mS and $W^{0}\approx 27$ nm, see Fig. 3(j). The experimental
energy gap is shown to scale as $E_{gap}(W)=\frac{\alpha }{W-W^{\ast }}$
with $\alpha =0.2$ eVnm and $W^{\ast }=16$ nm. Our fit gives the linear
scaling with $\alpha =2.1$ eVnm and $W^{\ast }=11$ nm (Fig. 3(k)). The
experimentally extracted width $W^{\ast }\approx W^{0}$ was interpreted as
an inactive edge width of GNR. The width of the disorder region in our
calculation is just one atomic row such that nominally inactive edge width
is just a fraction of a nanometer. However, as shown above, the
disorder-induced localization leads to the strong enhancement of the
electron density in the surface-like states not participating in the
transport. Their width on each side of the ribbon is $l_{loc}\sim 5-10$ nm
which is consistent with calculated values of $W^{\ast },W^{0}\sim 2l_{loc}$%
. We therefore speculate that both $W^{\ast }$ and $W^{0}$ can indeed be
interpreted as inactive edges, whose width is however determined by the
extent of the disorder induced localized surface-type states.

Our value for the energy gap is about of factor of $\sim 10$ larger than the
experimentally extracted one. One of the reasons for this difference can be
attributed to the phase breaking effects that would suppress localization of
electrons. Recent experiments indicate that the phase coherence length in
graphene $l_{\phi }\sim 3-5\mu $m@0.25K and $\sim 1\mu $m@1K. It is
therefore reasonably to expect that in the measurements of Han \textit{et al.%
} (performed at $T>1.7K$) $l_{\phi }$ is smaller than the device length ($%
L\sim 1$ $\mu $m), such that the inelastic processes may play an important
role in formation/suppression of the gap. In real ribbons the energy gap in
the conductance might be due both to modification of the electronic
structure caused by edges (as the DFT calculations show\cite%
{Cohen,Barone2006}) as well as due to Anderson-type localization as
discussed above. The later is expected to depend on the coherent length (and
thus to be temperature sensitive when $l_{\phi }<L$), whereas the former is
practically not affected by the temperature. Thus, experimental study of the
conduction gap in the mK range (i.e. exploring the transitions between the
regimes $l_{\phi }>L$ and $l_{\phi }<L)$ might shed more light on the origin
of the gap. Another factor that might strongly affect the gap formation is
the electron interaction. Indeed, because the LDOS is enhanced by many
orders of magnitude near the edge imperfections, it is reasonable to expect
that the Hartree potential would contribute significantly to the total
confining potential and thus affect the conduction gap. We therefore hope
that our results will motivate further studies of electron interaction and
phase breaking effects in realistic GNRs.

All the results presented above correspond to zero temperature. We also
performed calculations in the temperature range 0-200K which, as expected,
show gradual suppression of the gap as temperature raises. The energy
broadening at 200K is roughly the same as the energy gap for the 24nm-wide
ribbon and hence at this temperature the gap disappears due to the
temperature averaging.

\begin{figure}[tbh]
\includegraphics[scale=0.9]{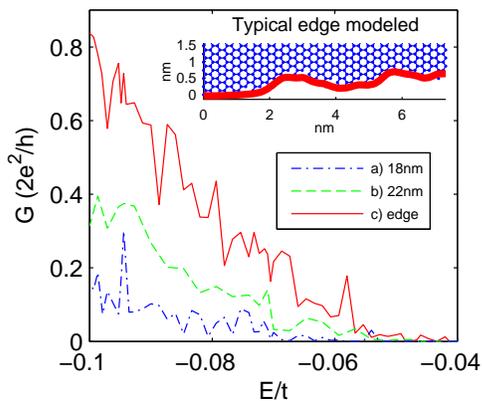}
\caption{(color online) Conductance for a $W=24$nm wide armchair GNR with extended edge
disorder as shown in the inset; $L=1\mu$m. The legend indicates the minimum width of the
ribbon. Each transmission curve is averaged over a set of edge contour configuration with
the same minimum width. Red solid curve is averaged transmission for a ribbon with
impurities in the outermost row only. Temperature $T=0$.} \label{fig:edge}
\end{figure}

In real samples fabricated by etching techniques a variation of the ribbon
width near the edges is expected to be much larger in comparison to the
model used above, of the order of at least several nanometers\cite%
{Han2007,Chen2007}. Figure \ref{fig:edge} shows the conductance of the armchair GNR with
the boundary modelled as a superposition of Lorentzians (see inset for illustration of a
typical edge). We focus on the 1$\mu$m long and 24nm wide armchair GNR with the edge of
the narrowest constriction of 18nm and 22nm. For the ribbons with the largest width
variation the overall conductance is somehow suppressed and the energy gap increases.
However, even though the conductance is apparently more strongly affected for the ribbons
with larger width variation, the conductance of the GNRs for different models of a
disordered edge is qualitatively very similar. We therefore expect that the utilized
model of the imperfect edge (with missing atoms in the outermost row) already captures
all the essential physics of realistic GNRs.

It is important to stress that in the striking contrast to the GNRs, the
Anderson-type localization near the edges is absent in conventional
heterostructure semiconductor wires because their edges are smooth on an
atomic scale. It should be also noted that the strong enhancement of the
LDOS near defects at the edges and formation of the surface-like state for
sufficiently high disorder concentration can be detected with the help of
STM \cite{Kobayashi}.

Recently an alternative explanation of the energy gap in GNR:s based on the
assumption of the Coulomb blockaded (CB) transport regime in GNR has been
suggested by Sols \textit{et al.} \cite{Sols2007}. While we do not challenge
their theory as such, our findings indicate that bare presence of a slightly
disordered edge (much weaker that it would be required for the CB regime) is
already sufficient to explain the gap formation.

Finally, we also performed conductance calculations studying the effect of
charged impurities which is believed to be the main mechanism of scattering
in the bulk graphene\cite{Ando,Hwang}. We model them by adding randomly the
onsite potential $V_{i}=\frac{e}{4\pi \epsilon _{0}\epsilon _{r}a_{cc}}$
\cite{Fisher}, where $\epsilon _{r}=5$ and $a_{cc}=$1.42\AA\ is the
carbon-carbon atom distance. For realistic impurity densities $n\sim 1\times
10^{16}$m$^{\text{-2}}$ and strength $V_{i}\sim $ 0.3eV \cite{Ando,Hwang} we
find that the conductance remains practically unaffected, which rules out
the charged impurities as the origin of the energy gap formation in the GNR.

\textit{Conclusion.} We study the effect of the edge disorder on the
conductance on the GNRs and find that even a very modest defect
concentration causes a strong Anderson type localization at the edges giving
rise to the conduction gap in accordance to recent experiments.
%


\begin{thebibliography}{99}
\bibitem{Novoselov2004} K.\ S.\ Novoselov, \textit{et al.}, Science, \textbf{%
306}, 666 (2004).

\bibitem{unstableGraphene}
L.\ D.\ Landau and E. M. Lifshitz, \textit{\ Statistical Physics, Part I}
(Pergamon, Oxford, 1980).

\bibitem{Novoselov2005_nature} K.\ S.\ Novoselov, \textit{et al.}, Nature,
\textbf{438}, 197 (2005).

\bibitem{Zhang} Y. Zhang, Y.-W. Tan, H. L.\ Stormer, P. Kim, Nature \textbf{%
438}, 201 (2005).

\bibitem{Schedin_nature} F. Schedin, \textit{et al.}, Nature materials
\textbf{6}, 652 (2008).

\bibitem{Novoselov2007_nature} K.\ S.\ Novoselov, A.\ K.\ Geim, Nature
Materials \textbf{6}, 183 (2007).

\bibitem{Chen2007} Z. Chen, Y.-M. Lin, M. J.\ Rooks and P. Avouris, Physica
E \textbf{40}, 228 (2007).

\bibitem{Fujita1996} M. Fujita, K. Wakabayashi, K. Nakada and K. Kusakabe,
J. of the Phys. Soc. of Japan \textbf{65}, 1920 (1996).

\bibitem{Nakada} K. Nakada, M. Fujita, G. Dresselhaus and M. S. Dresselhaus,
Phys. Rev. B \textbf{54}, 17954 (1996).

\bibitem{Han2007} M. Y.\ Han, B. \"{O}zyilmaz, Y. Zhang, and P. Kim, Phys.\
Rev.\ Lett.\ \textbf{98}, 206805 (2007).

\bibitem{Li2008_Science} X. Li, X. Wang, L. Zhang, S. Lee and H. Dai,
Science \textbf{319}, 1229 (2008).

\bibitem{Louis2007} E.\ Louis, J.\ A.\ Verg\'{e}s, F.\ Guinea, and G.\
Chiappe, Phys.\ Rev.\ B \textbf{75}, 085440 (2007).

\bibitem{Gunlycke} D. Gunlycke, D. A. Areshkin, and C. T. White, Appl. Phys.
Lett. \textbf{90}, 142104 (2007).

\bibitem{Li2008} T.\ C.\ Li and S.-P. Lu, Phys.\ Rev.\ B \textbf{77}, 085408
(2008).

\bibitem{Querlioz2008} D.\ Querlioz, Y.\ Apertet, A.\ Valentin, K.\ Huet,
A.\ Bournel, S.\ Galdin-Retailleau, and P.\ Dollfus, Appl.\ Phys. Lett.\
\textbf{92}, 042108 (2008).

\bibitem{Lherbier2008} A. Lherbier, B. Biel, Y.-M. Niquet, and S. Roche,
Phys.\ Rev.\ Lett.\thinspace\ \textbf{100}, 036803 (2008).

\bibitem{Cohen} Y.-W. Son, M. L.\ Cohen, and S. G.\ Louie, Phys.\ Rev.\
Lett. \textbf{97}, 216803 (2006); L. Yang, C.-H. Park, Y.-W. Son, M. L. Cohen, and S. G.
Louie, Phys.\ Rev.\ Lett. \textbf{99}, 186801 (2007).

\bibitem{Barone2006} V. Barone, O. Hod and G. E.\ Scuseria, Nano Lett.
\textbf{6}, 2748 (2006).

\bibitem{Xu2008} H. Xu, T.\ Heinzel, M.\ Evaldsson and I.\ V.\ Zozoulenko,
Phys. Rev. B, 2008, in press (arXiv:0804.0375v1 [cond-mat.mes-hall]).

\bibitem{Sols2007} F.\ Sols, F.\ Guinea, and A.\ H.\ Castro Neto, Phys.\
Rev.\ Lett. \textbf{99}, 166803 (2007).

\bibitem{Miao} F. Miao \textit{et al.}, Science \textbf{317}, 1530 (2007).

\bibitem{Russo} S. Russo \textit{et al.}, Phys. Rev. B \textbf{77}, 085413
(2008).

\bibitem{Ando} T. Ando, J. Phys. Soc. Japan \textbf{75}, 074716 (2006).

\bibitem{Hwang} E. H. Hwang, S. Adam, and S. Das Sarma, Phys. Rev. Lett.
\textbf{98}, 186806 (2007).

\bibitem{Fisher} J. Alicea and M. P. A. Fisher, Phys. Rev. B \textbf{74},
075422 (2006).

\bibitem{Kobayashi} Y. Kobayashi, K. I. Fukui, T. Enoki, K. Kusakabe, Y. Kaburagi,
 Phys. Rev. B 71, 193406
(2005).
\end{thebibliography}
\end{document}